\documentclass{aa}  
\usepackage{graphics,graphicx}
\usepackage{xcolor}
\usepackage{enumitem}
\usepackage{float}
\usepackage{caption}
\usepackage{subcaption}

\usepackage[utf8]{inputenc}
\DeclareUnicodeCharacter{202F}{\,} 

\usepackage{txfonts}
\usepackage{longtable}
\usepackage[unicode=true,pdfusetitle,
 bookmarks=true,bookmarksnumbered=false,bookmarksopen=false,
 breaklinks=false,pdfborder={0 0 0},pdfborderstyle={},backref=false,colorlinks=true]
 {hyperref}
 
\hypersetup{
 urlcolor=blue,citecolor=blue}

\begin{document}  
   \title{XRISM spectroscopy of a crowded Galactic center region - III.\\
   S, Ar and Ca ISM Absorption Features in the spectrum of MAXI J1744–294}  
   \author{E. Gatuzz\inst{1},   
           M. Parra\inst{2}, 
           M. F. Hasoglu\inst{3}, 
           T. W. Gorczyca\inst{4},
           S. Mandel\inst{5},
           K. Mori\inst{5},
           K. Matsunaga \inst{6},
           H. Uchiyama \inst{7},
           M. Nobukawa \inst{8}\and
           M. Shidatsu \inst{2}
          }  
          
   \institute{Max-Planck-Institut f\"ur extraterrestrische Physik, Gie{\ss}enbachstra{\ss}e 1, 85748 Garching, Germany\\
              \email{egatuzz@mpe.mpg.de}
         \and
         Department of Physics, Ehime University, 2-5, Bunkyocho, Matsuyama, Ehime 790-8577, Japan
         \and
             Department of Computer Engineering, Hasan Kalyoncu University, 27100 Sahinbey, Gaziantep, Turkey
         \and
              Department of Physics, Western Michigan University, Kalamazoo, MI 49008, USA     
         \and
              Columbia Astrophysics Laboratory, Columbia University, New York, NY 10027, USA
        \and 
            Department of Physics, Graduate School of Science, Kyoto University, Kitashirakawa Oiwake-cho, Sakyo-ku, Kyoto 606-8502, Japan
        \and 
            Faculty of Education, Shizuoka University, 836 Ohya, Suruga-ku, Shizuoka, Shizuoka 422-8529, Japan
        \and
            Faculty of Education, Nara University of Education, Nara, 630-8502, Japan
            } 
 
   \date{Received XXX; accepted YYY}  
  \abstract  
{We present a comprehensive study of X-ray absorption by sulfur (S), argon (Ar), and calcium (Ca) in the interstellar medium (ISM) along the line of sight to the low-mass X-ray binary MAXI~J1744$-$294, using high-resolution {\it XRISM} Resolve spectra complemented by {\it Chandra} HETG data. 
The analysis employs an updated {\tt ISMabs} model, incorporating newly computed $R$-matrix photoabsorption cross-sections for Ca\,{\sc i}--Ca\,{\sc iii}, and existing cross-sections for higher ionization states. 
We find that S and Ar are predominantly in low-ionization states, with S\,{\sc ii} and Ar\,{\sc ii} dominating the cold and warm ISM phases, while higher-ionization species are constrained by upper limits. 
Calcium is primarily detected in low-ionization states, consistent with strong depletion into dust grains, with only marginal contributions from highly ionized ions. 
Using the measured ionic column densities, we infer hydrogen column densities of $N_{\rm H} \sim 1.1$-$1.3 \times 10^{23}\,\mathrm{cm^{-2}}$ from S and Ar, while the Ca-based value, tracing the neutral ISM, is in agreement with these estimates, highlighting the consistency across different tracers. 
Our results demonstrate the diagnostic power of combining multiple elements to probe ISM ionization structure, elemental depletion, and dust composition, and provide the first X-ray constraints on calcium absorption in the interstellar medium.   
}

\titlerunning{ISM X-ray absorption towards MAXI J1744-294}
\authorrunning{Gatuzz et al.}
\maketitle
 
\section{Introduction} 
The interstellar medium (ISM) is a fundamental component of Galactic ecosystems, playing a central role in regulating star formation, stellar feedback, and Galactic chemical evolution \citep{won02,big08,ler08,lad10,lil13}. 
Composed of gas and dust distributed between stars, the ISM is structured into multiple phases spanning a wide range of physical conditions, with characteristic temperatures from $\sim10$ to $10^{6}$~K \citep[e.g.,][]{mck77,fal05,ton09,dra11,jen11,rup13,zhu16,sta18}. 
Understanding the composition, ionization structure, and depletion of elements across these phases is essential for constraining both small-scale astrochemical processes and large-scale Galactic dynamics.  

High-resolution X-ray spectroscopy offers a powerful and direct method to investigate the ISM by probing photoabsorption features imprinted on the spectra of bright Galactic X-ray sources.
In this ``X-ray background lamp'' technique, absorption edges and resonance lines arising from deep K- and L-shell electronic transitions provide diagnostics of elemental abundances, ionization states, and the presence of dust and molecules along the line of sight \citep{cos12,cos22}. 
Over the last decade, X-ray absorption studies of the ISM have focused primarily on the K-shell edges of abundant elements such as C \citep{gat18a}, O \citep{pin10,gat13a,gat13b,joa16,gat18b,gat19,psa20}, Ne \citep{gat14,gat15}, Mg \citep{cos12}, N \citep{gat21}, Si \citep{zee19,gat20,yan22}, S \citep{gat24b,psa24}, Ar \citep{gat24} and Fe \citep{pin13,gat16,rog18,wes19,rog21,psa23,corr24}, as well as the Fe L-shell region, yielding key insights into gas-dust partitioning and interstellar chemistry.  

The launch of {\it XRISM} has significantly advanced this field by delivering unprecedented spectral resolution with the Resolve microcalorimeter. 
This capability enables the detection and detailed modeling of narrow absorption features, allowing individual ionization states and potential solid-phase contributions to be disentangled with far greater fidelity than previously possible.
In this context, elements with relatively unexplored or poorly constrained X-ray absorption signatures --such as sulfur (S), argon (Ar), and calcium (Ca)-- provide new opportunities to deepen our understanding of the ISM. 

Sulfur is a key element in interstellar chemistry and astrobiology \citep{hux86,wal20,ols21}, yet its behavior in the ISM remains enigmatic. 
Observations indicate that sulfur is largely undepleted in diffuse ISM environments \citep{sof94,sav96b,mar02,how06,neu15}, whereas in dense and high-column-density regions it can be strongly depleted, with up to $\sim99\%$ of sulfur removed from the gas phase in molecular clouds \citep{jen09,woo15,laa19,hil22,fue23,psa24}. 
Only a minor fraction of this depleted sulfur is observed in known S-bearing molecules and ices, implying that the dominant sulfur reservoir resides in dust grains \citep{opp74,tie94,pal97,woo15,vas18,riv19,fue19}. 
Proposed solid-phase sulfur carriers include icy grain mantles, sulfur allotropes, refractory sulfur residues, and sulfur incorporated into silicate or metal-bearing grains, with the dominant form depending sensitively on environmental conditions such as cloud age, density, shocks, and grain composition \citep{smi91,cas94,cha97,hat98,vit01,vdt03,wak04,jim11,wak11,jim14,laa19,shi20,caz22,per24}.  

Argon, while chemically inert, is also an important tracer of ISM physical processes. 
Its abundance and ionization balance provide insights into shock heating, cosmic-ray ionization, and nucleosynthetic enrichment from supernovae \citep{ogl09,dop18,dop19}. 
Argon emission and absorption features have been used to probe star-forming regions and supernova remnants \citep{lop10,str23}, while its relative abundance with respect to other elements constrains metallicity and Galactic chemical evolution \citep{hua23}. 
However, argon depletion onto dust grains remains poorly constrained, partly due to uncertainties in stellar argon abundances and supernova yields \citep{ama21,san13,leu18,kob20,pal21}. 
High-resolution X-ray absorption studies provide a complementary avenue to assess the ionization state and potential depletion of argon in the ISM. 

Calcium represents a particularly compelling, yet largely unexplored, element in the context of X-ray ISM absorption studies. 
As a highly refractory element, calcium is expected to be heavily depleted from the gas phase and locked into dust grains in most ISM environments \citep{sav96,wil05,sar14}.  
The sensitivity and spectral resolution of XRISM now make it possible, for the first time, to investigate calcium X-ray absorption features and to assess their diagnostic potential for probing dust composition and extreme depletion in interstellar environments.

In this work, we present a comprehensive study of X-ray ISM absorption features associated with sulfur, argon, and calcium in the {\it XRISM}-Resolve spectrum of the black hole low-mass X-ray binary candidate MAXI~J1744$-$294/Swift J174540.2-290037, and complemented with other X-ray instruments.
This analysis combines the diagnostic power of these three elements to probe the multiphase ISM, elemental depletion, and dust composition along a Galactic line of sight, and includes the first X-ray absorption study of calcium in the ISM.
The paper is organized as follows.
Section~\ref{sec_data} describes the XRISM observations and data reduction.
Sections~\ref{sec_s_fit},\ref{sec_ar_fit} and \ref{sec_ca_fit} outline the absorption modeling of the Sulfur, Argon, and Calcium K-edges, respectively.
The results are presented and discussed in Section~\ref{sec_dis}.
Finally, our main conclusions are summarized in Section~\ref{sec_con}.

\section{Data reduction and continuum modeling }\label{sec_data} 
Here we describe the data reduction and continuum modeling.   

\subsection{{\it XRISM} observation}\label{subsec:xrism_reg}  
Following the discovery of its outburst by MAXI in January 2025 \citep{Kudo2025}, MAXI J1744-294 was subjected to an extensive multi-instrument campaign to study its outburst evolution, whose analysis is presented in detail in \citep{man26a}.
A {\it XRISM} observation was performed as a Director Discretionary Time (DDT) program on 2025 March 3, for a net exposure of 71~ks (ObsID~901002010), with Resolve operated in the OPEN filter configuration. The data reduction  of XRISM's microcalorimeter Resolve \citep{Ishisaki2022,Porter2024}, calibration procedures, and background modeling are described in detail in \textcolor{blue}{Parra et al. (submitted), hereafter paper I.}; here we summarize the main aspects relevant to the present analysis.

Because of {\it XRISM}'s modest angular resolution and the fixed pixel geometry of Resolve, spatial-spectral mixing between MAXI~J1744$-$294, nearby point sources (including the Neutron Star AX J1745.6-2901), and diffuse Galactic Center emission is unavoidable. 
To quantify its impact, in paper I, we defined two alternative Resolve extraction regions: a ``large'' region maximizing the enclosed PSF fraction and signal-to-noise ratio, and a ``small'' region restricted to the brightest pixels to minimize contamination. 
Diffuse emission was estimated using a previous XRISM observation with a similar field of view, accounting for small pointing differences between the datasets, while the nearby neutron-star source AX~J1745.6$-$2901 was modeled separately, and fitted jointly during spectral analysis. 
Known problematic pixels were excluded from the analysis. In this paper, we exclusively use the products generated from the ``large'' region, whose empirical background modeling is more accurate at low energies compared to the physically motivated models used for the small region, and tailored to the highly ionized emission lines. Moreover, the larger angular coverage of this region makes it less subject to uncertainties due to the modeling of the point-spread-function of Resolve, and less affected by the Dust Scattering Halo (DSH) surrounding MAXI J1744-294.

The Resolve energy scale was calibrated on a per-pixel basis using the onboard $^{55}$Fe calibration source, with corrections applied for pixel temperature evolution. 
One pixel affected by gain instabilities was excluded, and a recalculated gain solution was applied to address insufficient calibration photons in the brightest pixel. 
Spectra were extracted from both high-resolution (Hp) and medium-resolution (Mp) primary events following updated calibrations. 
Spectral responses were generated using combined rmfs in the extra-large (X) format generated from the 2-12keV event lists, along with arfs computed with point-source distribution for MAXI J1744-294 and AX J1745.6-2901, and a flat distribution larger than the Resolve field of view for the background.
Systematic uncertainties related to recent {\it XRISM} calibration updates were evaluated, and the calibration version providing the best internal consistency across extraction regions (CALDB11) was adopted.
Additional bright-source effects on the Resolve energy scale and spectral resolution were assessed and found to be small compared to the statistical uncertainties, and therefore do not qualitatively affect our results. 
 
\subsection{{\it Chandra} observations}
{\it Chandra} observed MAXI~J1744$-$294 on 2025 March 9 with the ACIS-S instrument, operated in 1/8th subarray mode (S2+S3) and combined with the High-Energy Transmission Grating (HETG) to reduce pile-up \citep{man26a}. 
The source was detected as a very bright, heavily piled-up object near the center of the ACIS-S field, with clearly visible HEG and MEG dispersion arms whose intersection marks the zeroth-order position \citep{man25b,man26a}. 
We determined the source coordinates using the CIAO task {\tt tg\_findzo}, which locates the zeroth-order position from the grating dispersion geometry. 
This position was independently confirmed using the {\tt celldetect} algorithm, which provides a positional uncertainty of 1.2\arcsec\ (68\% confidence). 
The derived {\it Chandra} position is consistent with the radio position reported by MeerKAT and lies at a projected separation of $<$20\arcsec\ from Sgr A*.  
The data were processed using the {\tt chandra\_repro} script for implementing {\it CIAO} commands. 
The argon and calcium X-ray absorption edges are accessible with the HETG exclusively via the High Energy Gratings (HEG), which is therefore the spectrum analyzed in the following.

\subsection{Continuum modeling}  
For the continuum modeling, we follow the methodology described in \textcolor{blue}{Parra et al. (submitted) b, hereafter paper II}. 
The X-ray continuum is modeled simultaneously for the black hole MAXI~J1744$-$294, the nearby neutron star AX~J1745.6$-$2901, and the diffuse Galactic Center emission, as these components significantly overlap spatially and spectrally in the Resolve field of view and must be treated self-consistently. 
For the black hole, the continuum consists of a multicolor disk blackbody component describing the thermal accretion disk, Comptonized emission ({\tt thcomp}) accounting for the hard non-thermal tail, modified by interstellar absorption. 
The neutron star spectrum was modeled with a combination of a disk blackbody and a higher temperature thermal blackbody , also subject to strong absorption. While it contributes substantially to the total flux at energies above $\sim$6--7~keV, its contribution compared to MAXI J1744-294 remains negligible at low energies. We note that MAXI J1744-294 exhibits strong emission lines (see paper II), and AX J1745.6-2901 strong absorption lines (Matsunaga et al. in prep.), but these features are only present for highly ionized lines and thus do not affect any of the results presented in this analysis. 
The diffuse emission is included as a uniform background component with fixed parameters directly imported from the fits of the same region in the archival XRISM observation, as described in paper I. 
Several continuum parameters that are weakly constrained at high spectral resolution are fixed to the best-fit values obtained from joint {\it XMM-Newton}, {\it NuSTAR}, and {\it XRISM} broadband fits, themselves detailed in \citep{man26a}, ensuring a stable continuum against which narrow ISM features can be robustly characterized.
In the following, all spectral fitting was done with the {\sc xspec} package (version 12.15.1\footnote{http://heasarc.nasa.gov/xanadu/xspec/}), using Cash statistics \citep{cas79}.  
We adopt the protosolar abundance units of \citet{lod09}.
Errors are quoted at 1$\sigma$ confidence level unless otherwise stated.

\section{Spectral fit of the S~K-edge}\label{sec_s_fit}  
We used a modified version of the {\tt ISMabs} model \citep{gat15}, which incorporates the photoabsorption cross-sections of all sulfur ions. 
Following \citet{gat24b}, our implementation includes S\,{\sc i}, S\,{\sc ii}, S\,{\sc iii}, and S\,{\sc xiv}, and the spectrum was fitted over the $2.42$--$2.51$~keV energy range. 
In this approach, the column densities of the relevant ionic species were treated as free parameters in the spectral fitting. 

\begin{figure}    
\centering 
\includegraphics[width=0.49\textwidth]{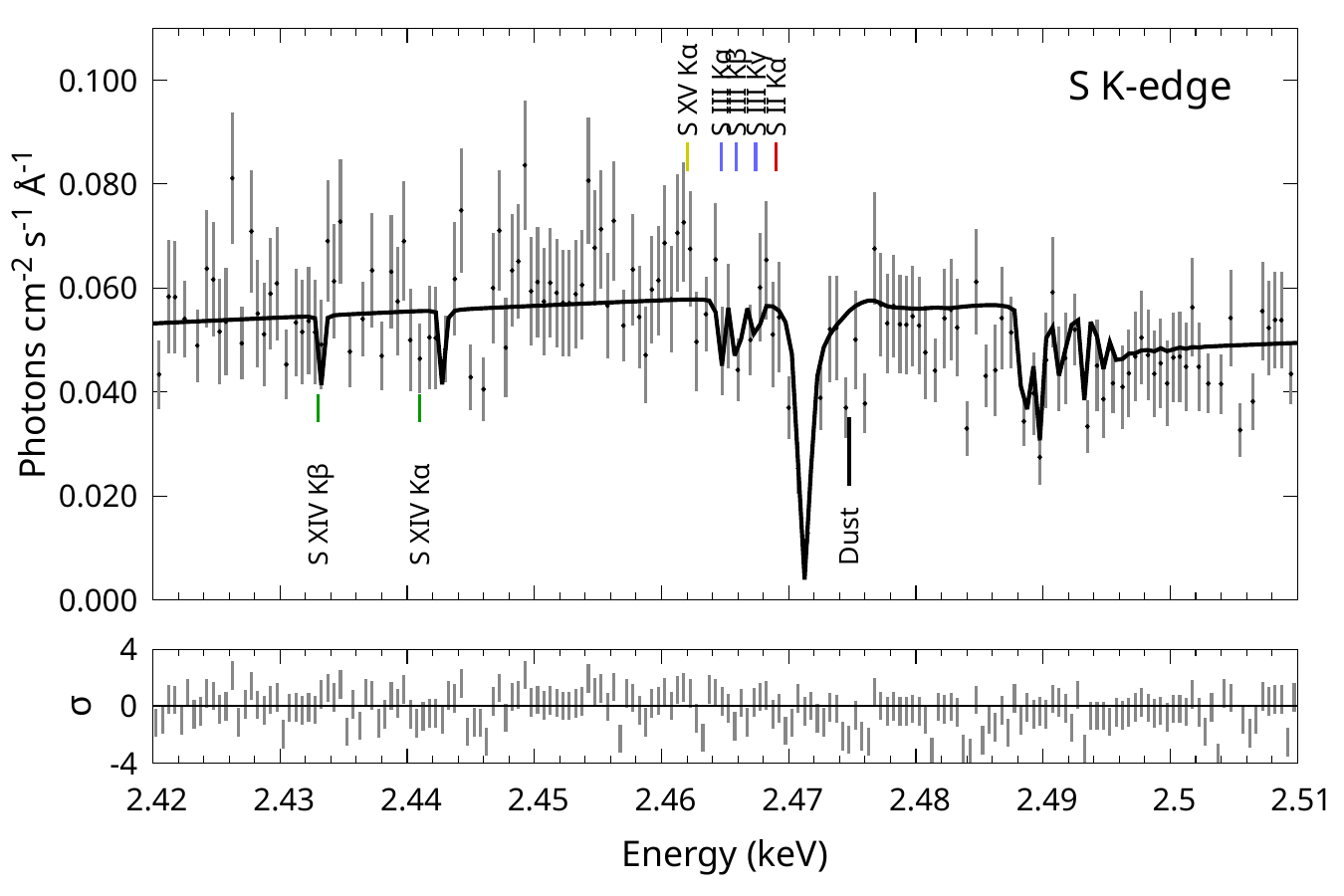} 
\caption{Best-fit results in the S~K-edge photoabsorption region for MAXI~J1744$-$294. 
Black points correspond to the observed flux, while the red curve shows the best-fit model. 
Residuals are displayed in units of $(data-model)/error$. 
The data has been rebinned for illustrative purposes.
The positions of the K$\alpha$ absorption lines are indicated for each ion. 
The S\,{\sc ii} line was shifted by approximately $\Delta E = -8.64$~eV \citep[similar to][]{cor25}.}
\label{fig_sulfur_spectra} 
\end{figure}

Figure~\ref{fig_sulfur_spectra} shows the resulting best-fit model. 
Black points represent the observed flux, while the red curve denotes the best-fit spectrum. 
Residuals are shown in units of $\sigma = (data - model)/error$, and the positions of the gaseous K$\alpha$ absorption lines are indicated for each ion. 
Parameter uncertainties were explored using Markov-chain Monte Carlo (MCMC) sampling with the Goodman-Weare algorithm in {\sc xspec}, employing 25 walkers and $2\times10^{6}$ steps, discarding the first $5\times10^{5}$ to ensure convergence. 
This procedure provided robust estimates of bulk velocities and turbulent line widths. 
The best-fit parameters are reported in Table~\ref{tab_ismabs_sulfur}.  

Following \citet{cor25} we fitted the shift of the S\,{\sc ii} K$\alpha$ line and found a value of $\Delta E  -8.64^{+0.33}_{-0.22}$~eV with a $\Delta cstat = 51$.
To test that such a shift is statistically warranted, we used a simulation-based $\Delta cstat$ method following \citet{buc23}.
For each case, 2000 spectra were simulated from the simpler model (i.e., without shift), and the 99th percentile of the resulting $\Delta cstat$ distribution defined the critical threshold ($\Delta cstat_{\rm crit}$) corresponding to a 1$\%$ false-positive rate.
We found $\Delta cstat_{\rm crit}=2.10$.
The best-fit statistic shows an improvement exceeding this threshold, indicating that the shift parameter is statistically warranted.
This supports the atomic data benchmarking proposed by \citet{cor25}.

\begin{table}
\caption{Best-fit sulfur column densities. \label{tab_ismabs_sulfur}}
\centering
\begin{tabular}{lc}
\hline  
Ion  &  Value \\
\hline
\hline 
S\,{\sc i}     & $<1$ \\
S\,{\sc ii}    & $186 \pm 22$ \\
S\,{\sc iii}   & $15_{-11}^{+14}$ \\
S\,{\sc xiv}   & $<20$ \\
S\,{\sc xv}    & $<1$ \\
S\,{\sc xvi}   & $<2$ \\
Alabandite     & $<69$ \\
Pyrrohtite     & $<78$ \\
Troilite       & $<68$ \\
$c$-stat/d.o.f.& $831/831$ \\
\hline 
\multicolumn{2}{l}{Column densities are in units of $10^{16}$~cm$^{-2}$.}
\end{tabular}
\end{table}

\section{Spectral fit of the Ar~K-edge}\label{sec_ar_fit} 
\begin{figure}    
\centering 
\includegraphics[width=0.49\textwidth]{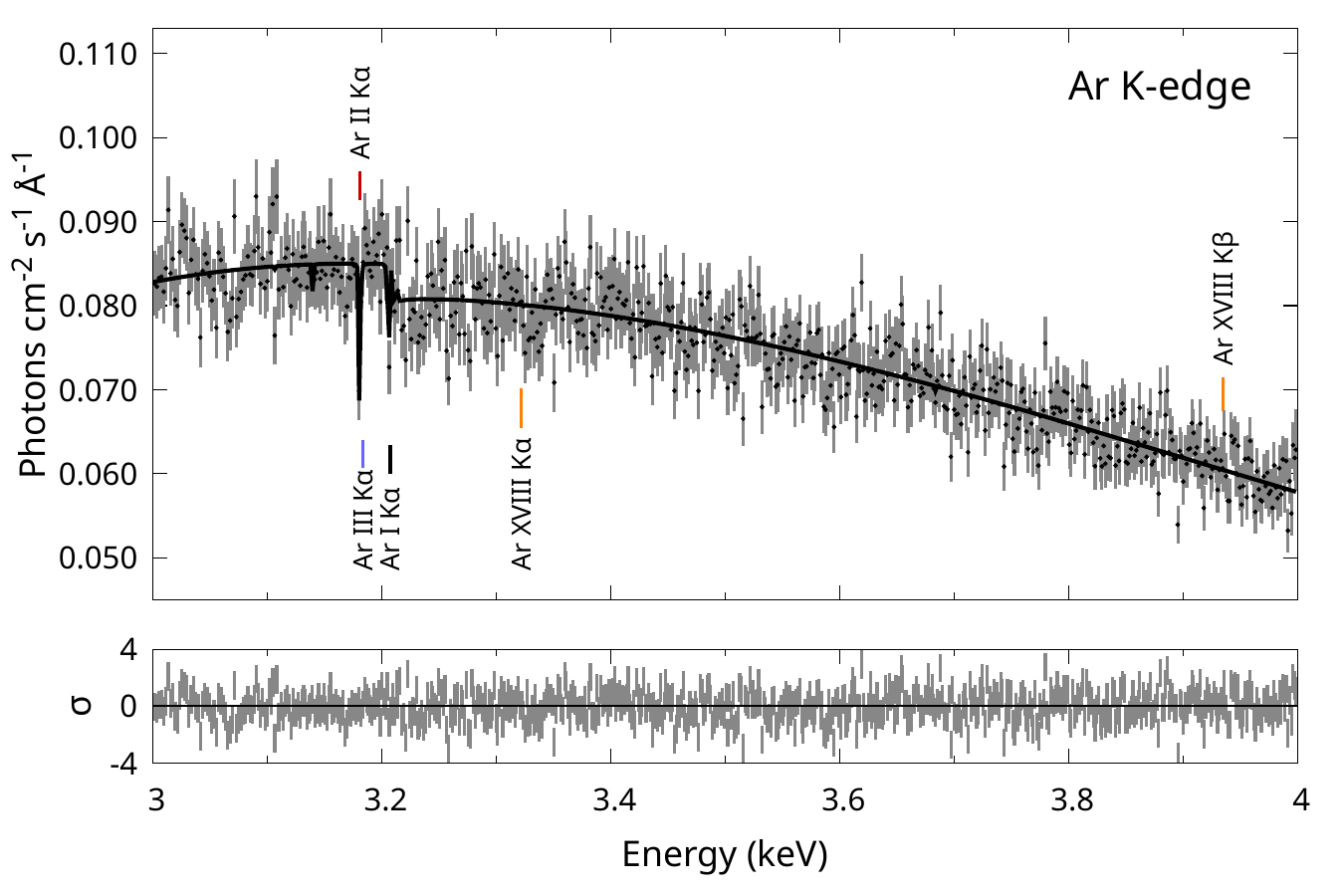} 
\caption{Best-fit results in the Ar~K-edge region for MAXI~J1744$-$294. 
Black points show the observed flux, and the red curve represents the best-fit model. 
Residuals are displayed in units of $(data-model)/error$, with K$\alpha$ absorption lines indicated for each ion.
The data has been rebinned for illustrative purposes.}
\label{fig_argon_spectra} 
\end{figure}

The Ar~K-edge was modeled using a modified version of the {\tt ISMabs} model \citep{gat15}, incorporating updated photoabsorption cross-sections for Ar\,{\sc i}, Ar\,{\sc ii}, Ar\,{\sc iii}, Ar\,{\sc xvi}, Ar\,{\sc xvii}, and Ar\,{\sc xviii} \citep{gat24c}. 
The best-fit model is shown in Figure~\ref{fig_argon_spectra}, where black points correspond to the observed flux and the red curve represents the model. 
Residuals are plotted in units of $\sigma = (data - model)/error$, and the positions of the K$\alpha$ absorption lines are indicated for each ion. 
Parameter estimation was performed using an MCMC approach, and the resulting column densities are reported in Table~\ref{tab_ismabs_argon}. 
The XRISM values are consistent with the upper limits obtained from {\it Chandra} data. 
During benchmarking of the atomic data, we found that any potential redshift of the Ar ions is constrained to $\Delta E < 2$~eV.
The \citet{cor25} statistical test shows a $\Delta cstat_{\rm crit}=6.60$, while the statistical change in the observation when including the redshift is $\Delta stat < 1$, thus the shift parameter is not statistically warranted.
Therefore, no shifts were applied to the Ar photoabsorption cross-sections. 

  \begin{table}
\caption{
\label{tab_ismabs_argon}
Best-fit argon column densities obtained.
 }
\centering
\begin{tabular}{lcc}
\hline  
Ion  &  {\it XRISM} Value &  {\it Chandra} Value\\
 \hline
\hline 
Ar\,{\sc i}& $<22.5$ & $<32$ \\
Ar\,{\sc ii}& $48\pm 12$ & $<35$ \\
Ar\,{\sc iii}& $<9.1$ & $<23$ \\
Ar\,{\sc xvi}& $< 1.2$ & $<45$ \\
Ar\,{\sc xvii}& $<0.9$ & $<37$ \\
Ar\,{\sc xviiii}& $0.52^{+0.76}_{-0.38}$ & $<38$ \\
$c-stat$/d.of.& $2164/2093$ & $284/413$ \\
\hline 
\multicolumn{2}{l}{ Column densities in units of $10^{16}$cm$^{-2}$ .}
 \end{tabular}
\end{table}

\section{Spectral fit of the Ca~K-edge }\label{sec_ca_fit} 

\subsection{Ca~K-edge photoabsorption cross-sections}\label{sec_ca_cross} 
To model the high-resolution Ca~K-edge observed with {\it XRISM}, we computed new photoabsorption cross-sections for Ca ions using the $R$-matrix method. 
The $R$-matrix approach provides a reliable and effective framework for theoretical computation of K-shell photoionization cross sections, incorporating important electron correlation, orbital relaxation, channel coupling, and Auger broadening effects. 
In this approach, configuration space is divided into two regions: an inner region, $r\le a$, where all the target $N$-electron probability density is located, and therefore all inter-electron interactions occur, and an outer region, $r>a$, where the ejected photoelectron (or valence electron, for the initial $N+1$-electron bound state)  moves primarily under the long-range potential of the residual $N$-electron ion~\citep{burke}, which for photoionization of neutral and positively-charged ions gives linear combinations of {\em analytic} regular and irregular Coulomb functions.
Thus, for, say, photoionization of neutral calcium, the $R$-matrix method allows for the consistent treatment of the initial $N=22$-electron Ca bound state and the final ionized Ca$^+$ $N=21$-electron state plus an outgoing photoelectron. 

In the inner region, the total wavefunction is expanded in a configuration-interaction (CI) basis that includes couplings between the $N$-electron target states and the outer electron (which is bound for the initial state and free for the final state). 
Electron exchange and correlation effects are explicitly incorporated within a finite boundary radius $r \le a$, and the computed $R$-matrix, which involves the eigenvalues and eigenvectors of the Hamiltonian, is analytically evaluated at the boundary $r=a$.
In the outer region, the photoelectron moves under the potential of the residual ion with negligible exchange effects. 
For these calculations, the long-range channel coupling was determined to be negligible, and thus the photoelectron wavefunctions are linear combinations of the regular and irregular Coulomb functions.

\begin{figure} 
\centering
\includegraphics[width=0.46\textwidth]{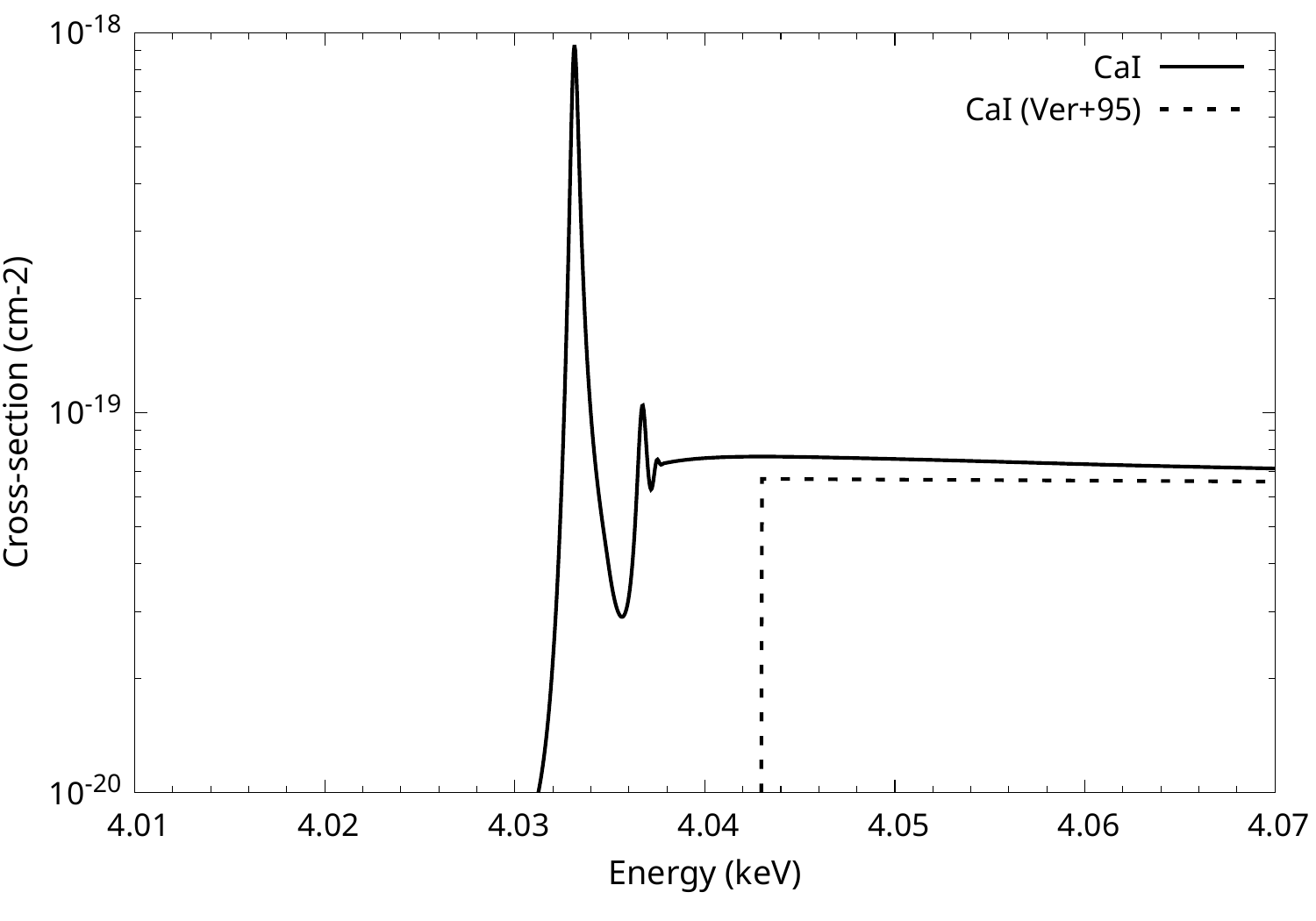}\\
\includegraphics[width=0.46\textwidth]{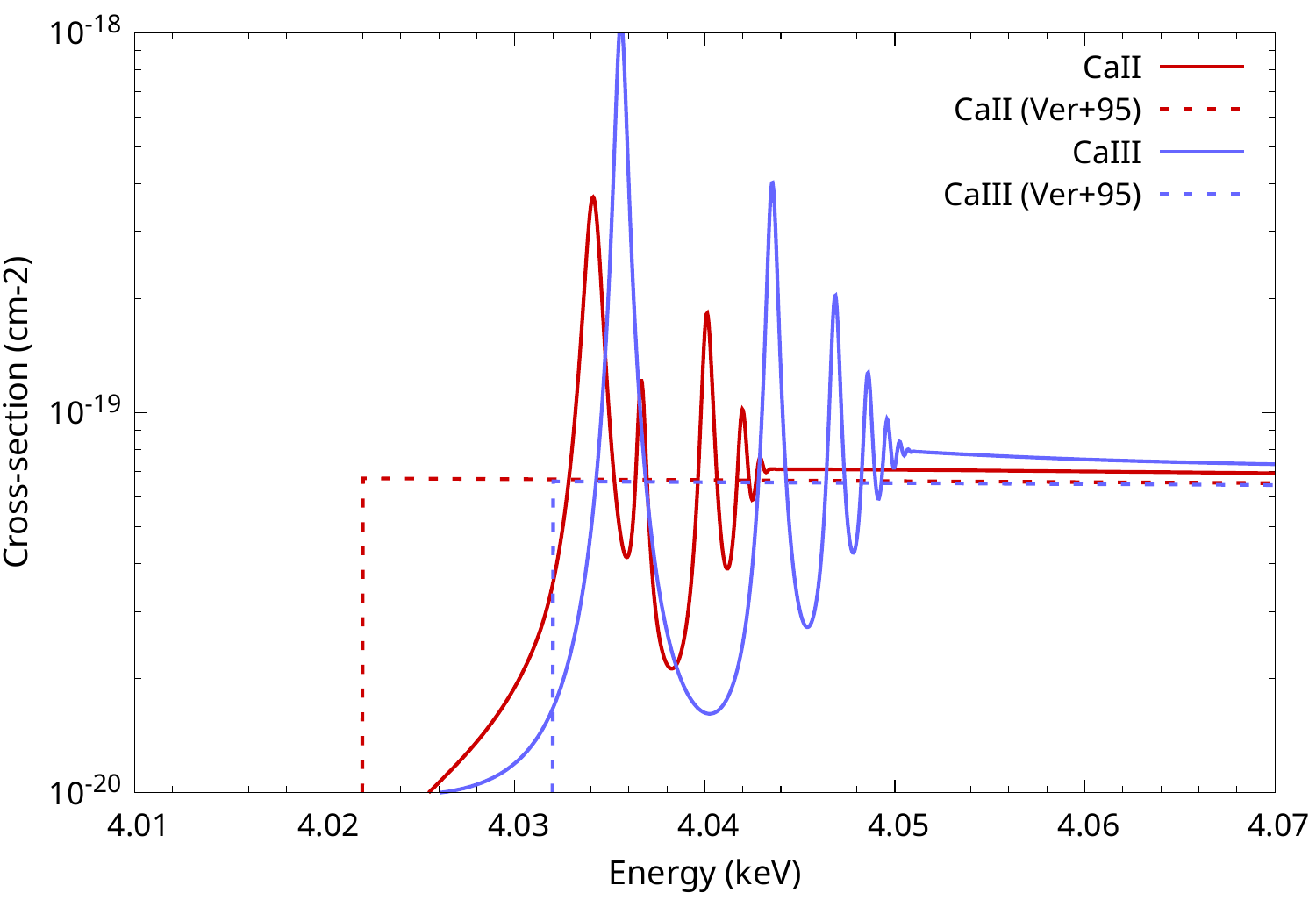}\\
\includegraphics[width=0.46\textwidth]{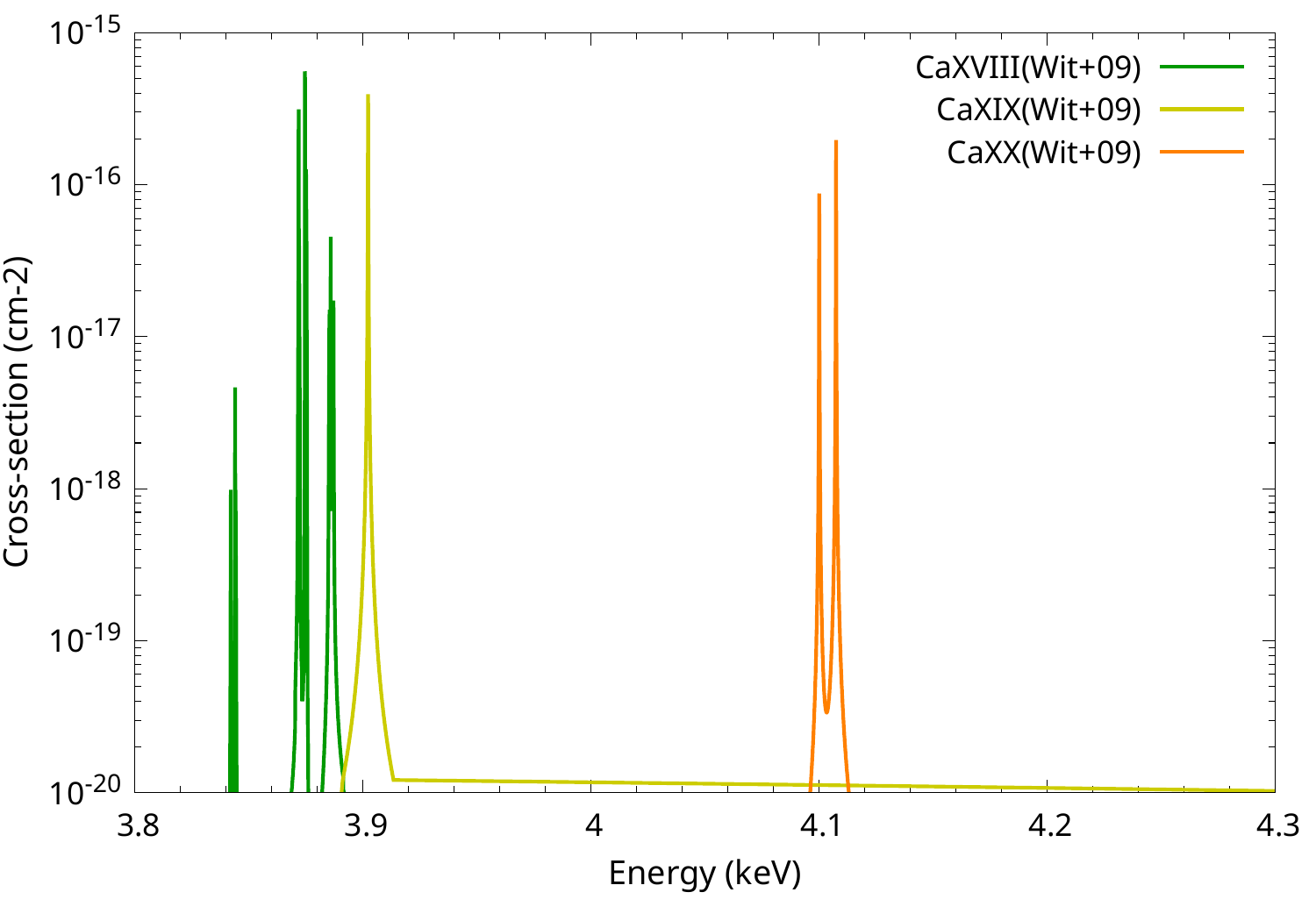}
\caption{Photoabsorption cross sections included in the model for Ca\,{\sc i} (top panel), Ca\,{\sc ii}, Ca\,{\sc iii} (middle panel), and Ca\,{\sc xviii}, Ca\,{\sc xix} and Ca\,{\sc xx} (bottom panel). 
For the last ones, our model includes the photoabsorption cross-sections from \citet{wit09}. 
For comparison, Ca\,{\sc i}-Ca\,{\sc iii} photabsorption cross-sections computed by \citet{ver95} are also included.
The overall profiles are alike, but the K-edge positions vary between the two calculations because this work accounts for orbital-relaxation effects when generating a basis set designed for inner-shell ionization.}\label{fig_ca_cross}
\end{figure}

Specifically for K-shell photoionization, which shows many strong inner-shell vacancy Rydberg series of resonances, the $R$-matrix method naturally accounts for resonance structures and Auger decay channels arising from $1s-$vacancy formation, allowing simultaneous treatment of direct and resonant processes. 
In the present paper, the so-called Belfast $R$-matrix codes \citep{ber95} were modified to include spectator Auger broadening effects via an optical potential~\citet{gor99}, and also to apply a pseudoresonance elimination method necessary when using pseudoorbitals and large configuration expansions~\citet{pseudo}. 
The $N$-electron target states consisted of linear expansions of Slater determinants of orbitals that were constructed from Hartree-Fock (HF) and multiconfiguration  Hartree-Fock (MCHF) methods~\citet{ffmchf}. 
Specifically, the $\{1s,2s,2p,3s,3p,3d,4s\}$ orbitals were computed as physical orbitals determined from Hartree-Fock calculations for the ground and excited states of the final ionized target state.   
Additional $\overline{4p}$, $\overline{3d}$, and $\overline{5s}$  pseudo-orbitals were generated  from MCHF calculations on the $1s$-vacancy states using all single and  double promotion configurations from the $1s-$vacancy target state configuration. 
The outer electron coupled to the $N$-electron target states is spanned by an additional basis of 100 orbitals that have zero derivative at $r=a$ (though not necessarily with zero magnitude, which is needed for spanning the photoelectron continuum state). 

Spectator Auger broadening effects on resonances below the Ca$^+$ K-shell threshold are incorporated by using an optical potential term $-i\Gamma/2$~\citep{pseudo} for the closed channels, where the Auger widths $\Gamma$ for the K-vacancy Ca$^+$ target states are
computed by performing an R-matrix calculation on electrons scattering from Ca$^{2+}$ and analyzing the Time Delay matrix~\citep{smi60}. 
The $R$-matrix formalism for the $e^-+1s^2n\ell^{q-2}$ scattering problem was applied with the use of consistent basis sets and configuration lists as for the $(q-1)$-electron scattering. 
Further details on the significance of spectator Auger broadening and its implementation with the $R$-matrix and Wigner Time Delay methods are discussed in previous works \citep{gor99, gor13}.

This theoretical method provides a robust foundation for astrophysical spectroscopic analysis, producing reliable K-shell cross sections and resonance structures essential for modeling X-ray absorption and emission in astrophysical plasmas, and was previously used for the identification of various lines and interpretation of high-resolution spectroscopic observations for C, Mg, Si, S, Ar ions \citep{has10,has14,gat20,gat24b,gat24}.  
As a consistency check on the present calculated results, we have compared the cross sections using both the length and velocity forms of the dipole operator, which analytically should be identical for exact wavefunctions, and found that the agreement is nearly perfect, differing by only a couple of percent.

Figure~\ref{fig_ca_cross} shows the resulting cross-sections for Ca\,{\sc i} (top panel), Ca\,{\sc ii}, Ca\,{\sc iii} (middle panel), and Ca\,{\sc xviii}, Ca\,{\sc xix}, Ca\,{\sc xx} (bottom panel, from \citealt{wit09}). 
Our calculations for Ca\,{\sc i}-Ca\,{\sc iii} are also compared to \citet{ver95}, showing overall similar profiles but shifted K-edge positions due to inclusion of orbital relaxation effects. 
These cross-sections were implemented in a modified {\tt ISMabs} model \citep{gat15}, allowing the column densities of each ion to be treated as free parameters in spectral fitting.

\subsection{Spectral fit results}\label{sec_ca_fit_results}   
\begin{figure}    
\centering 
\includegraphics[width=0.49\textwidth]{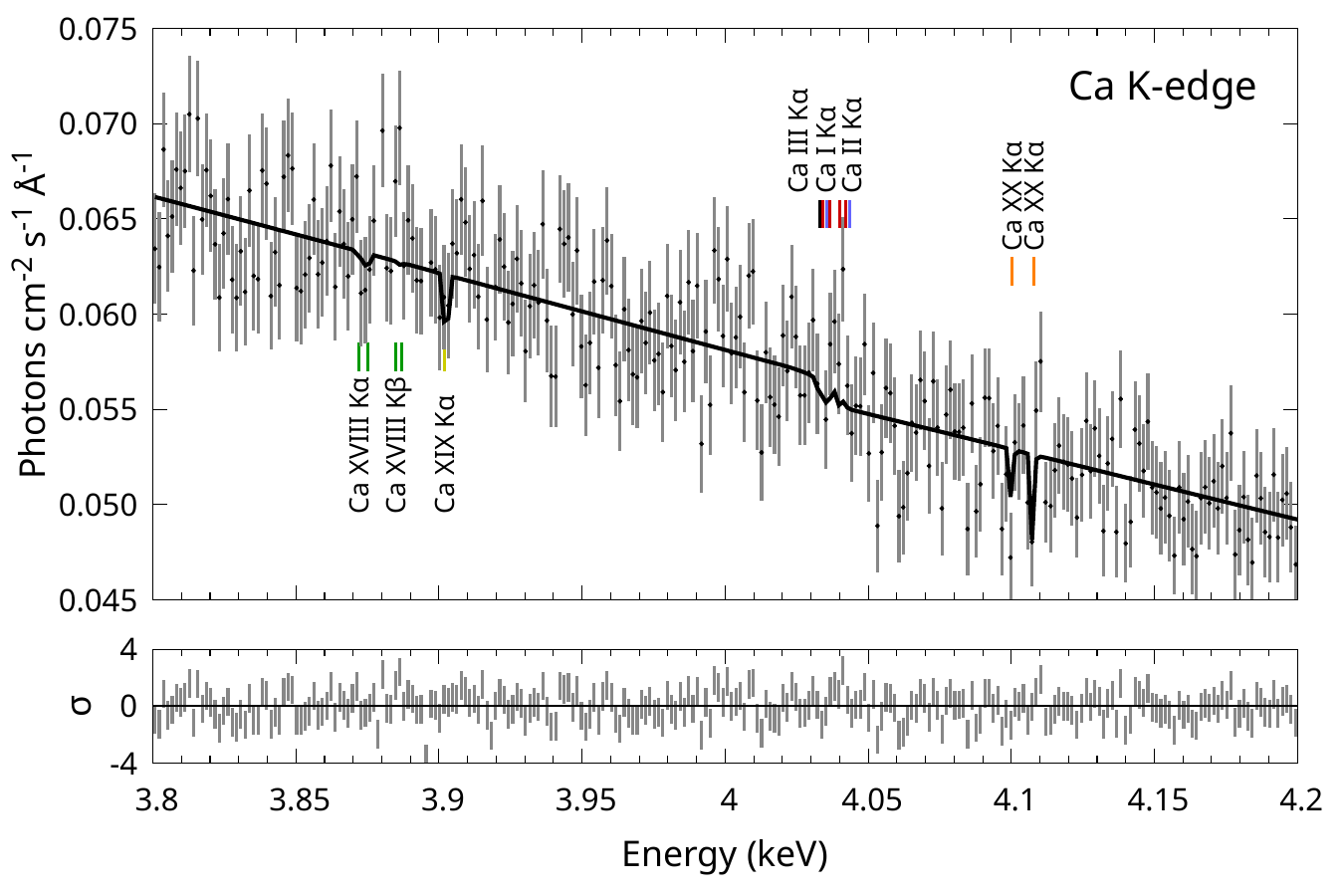} 
\caption{  Best-fit results in the Ca K-edge photoabsorption region for the LMXB MAXI~J1744-294. 
  Black points correspond to the observation in flux units, while the black line corresponds to the best-fit model. 
  Residuals are included in units of $(data-model)/error$. 
The data has been rebinned for illustrative purposes.
  The positions of the K$\alpha$ absorption lines are indicated for each ion, following the color code used in Figure~\ref{fig_ca_cross}.  
} \label{fig_calcium_spectra} 
\end{figure}

The Ca~K-edge in the MAXI~J1744$-$294 spectrum was modeled using the updated {\tt ISMabs} implementation described above. 
Figure~\ref{fig_calcium_spectra} shows the observed flux (black points) and best-fit model (red curve), with residuals in units of $\sigma = (data - model)/error$. 
K$\alpha$ absorption features for each ion are indicated. 
We performed MCMC sampling to explore the parameter space and obtain robust uncertainties. 
The resulting column densities are summarized in Table~\ref{tab_ismabs_calcium}, including upper limits derived from {\it Chandra} data. 
Our {\it Chandra} measurements are consistent with the XRISM results. 
The redshifts of the Ca ions were constrained to $\Delta E < 1$~eV, and the \citet{cor25} statistical test shows a $\Delta cstat_{\rm crit}=5.76$, while the statistical change in the observation when including the redshift is $\Delta stat < 1$.
Thus, the shift parameter is not statistically warranted, and no shifts were applied to the Ca photoabsorption cross-sections.

\begin{table}
\caption{
\label{tab_ismabs_calcium}
Best-fit calcium column densities obtained.
 }
\centering
\begin{tabular}{lcc}
\hline  
Ion  &  {\it XRISM} Value &  {\it Chandra} Value\\
 \hline
\hline 
Ca\,{\sc i}& $13.4^{+14.1}_{-9.5}$ & $ < 41$ \\
Ca\,{\sc ii}& $14.4^{+14.3}_{-10.2}$ & $ < 29$ \\
Ca\,{\sc iii}& $9.4^{+11.4}_{-6.8}$ & $ <38$ \\
Ca\,{\sc xviii}& $<0.3$ & $<10$ \\
Ca\,{\sc xix}& $<0.6$ & $<20$ \\
Ca\,{\sc xx}& $0.77_{-0.72}^{+1.34}$& $<15$ \\
$c-stat$/d.of.& $1065/999$ & $71/124$ \\ 
\hline 
\multicolumn{2}{l}{ Column densities in units of $10^{16}$cm$^{-2}$ .}
 \end{tabular} 
\end{table}

\section{Discussion}\label{sec_dis}
Here we discuss the main results obtained for each species analyzed.

\subsection{Sulfur: Comparison with Previous ISM Studies}
The best-fit sulfur column densities are broadly consistent with the upper limits reported by \citet{gat24b} for sources at similar Galactic longitudes and latitudes, with the cold ISM component dominated by S\,{\sc ii} and the hot component dominated by S\,{\sc xiv}. 
Using the upper limit on the sulfur dust component, we constrain the fraction of sulfur locked into dust to $<30\%$, corresponding to a dust-to-gas ratio $<0.42$. 
Given the large uncertainty on the S\,{\sc ii} column, this should be regarded as a conservative estimate. 
Previous studies report similar values: \citet{jen09} estimate up to 30\% depletion along lines of sight with $\log N_{\rm H} > 19.5$, \citet{psa24} found 40\% depletion toward Cygnus~X-2, and \citet{cor25} reported $40\% \pm 15\%$ for 4U~1630-472 and GX~340+0 using {\it XRISM}. 

Similar to \citet{cor25}, we compare the S\,{\sc ii} column density to the expected value from the hydrogen column density ($N_{\rm H}$).
The inferred hydrogen column density from 21\,cm observations is $N_{\rm HI} \approx 2 \times 10^{22}\,\mathrm{cm^{-2}}$ \citep{kal05}, while the source is also associated with a molecular hydrogen column density, thus providing a $N_{\rm H}=N_{\rm H_{gas}}+N_{\rm H_{2}}$ value of $1.34\times 10^{22}$ cm$^{-2}$ \citep{wil13}. 
However, these radio measurements are affected by saturation effects \citep{dic90,kal09}, providing formal lower limits on the total hydrogen column. 
From the S\,{\sc ii} column density, we infer $N_{\rm H} = 1.14\pm0.13\times10^{23}$~cm$^{-2}$, somewhat lower than the value derived by \citet{man26a} using {\it Swift} and {\it XMM-Newton}. 
X-ray-derived $N_{\rm H}$ values depend on the adopted abundance table, dust treatment, instrument resolution, and continuum model \citep{wil00,smi16,cor16}, which can explain differences with radio-based estimates. 

\subsection{Argon: Comparison with Previous ISM Studies}   
The argon column densities measured in this work are dominated by the singly ionized component, with \(N(\mathrm{Ar\,II}) = (4.8\pm1.2)\times10^{17}\,\mathrm{cm^{-2}}\), while neutral Ar and higher ionization stages are constrained only by upper limits (Ar I \(<2.25\times10^{17}\,\mathrm{cm^{-2}}\), Ar III \(<9.1\times10^{16}\,\mathrm{cm^{-2}}\), Ar XVI \(<1.2\times10^{16}\,\mathrm{cm^{-2}}\), and Ar XVII \(<9\times10^{15}\,\mathrm{cm^{-2}}\)), and a marginal detection of Ar XVIII \((5^{+7.6}_{-3.8})\times10^{15}\,\mathrm{cm^{-2}}\).
These results broadly agree with previous X-ray absorption studies, which also found that argon in the ISM is primarily in low ionization states along Galactic sightlines, with Ar II generally dominating the gas-phase abundance and only weak upper limits on highly ionized species \citep{gat24c}. 
Optical and ultraviolet absorption studies similarly report that neutral and singly ionized argon are the principal contributors to the total argon column in the diffuse ISM, with Ar I and Ar II absorption identified toward early-type stars and along diverse lines of sight \citep[e.g.,][]{sof04, jen09, erz25}. 
The relatively low columns of higher ionization stages, together with upper limits on Ar III and the hot-phase ions, are consistent with a predominantly neutral-to-warm ISM, and approximate mass fractions of $\sim90\%$, $\sim8\%$, and $\sim2\%$ in the neutral, warm, and hot phases, respectively \citep[e.g.,][]{yao06,pin13,gat18a}. 
Variations in argon depletion into dust, particularly toward the Galactic center where column densities are highest \citep{rob03,kalb09,gat24b}, may further suppress gas-phase absorption features and lead to underestimates of the argon abundance \citep{cos22}.  
The hydrogen column derived from Ar\,{\sc ii}, $N_{\rm H} = 1.34\pm0.32\times10^{23}$~cm$^{-2}$, is in agreement with that inferred from sulfur, reinforcing the consistency of these measurements across elements.

\subsection{Calcium: First Detection of X-ray Absorption in the ISM}  
Calcium column densities are dominated by low-ionization species (Ca\,\textsc{i}--Ca\,\textsc{iii}), with $N(\mathrm{Ca\,I}) = 1.34\times10^{17}$, $N(\mathrm{Ca\,II}) = 1.44\times10^{17}$, and $N(\mathrm{Ca\,III}) = 9.4\times10^{16}$~cm$^{-2}$, although the detection of the low-ionization elements are marginal and correspond mainly to upper limits. 
Only weak constraints are obtained for highly ionized calcium, with upper limits on Ca\,\textsc{xviii} and Ca\,\textsc{xix}, and a marginal detection of Ca\,\textsc{xx}. 
This represents the first X-ray study of calcium in the ISM, enabled by newly computed photoabsorption cross sections, so direct comparison with previous X-ray results is not possible. 

Optical and ultraviolet observations show that calcium is among the most heavily depleted elements, with gas-phase abundances suppressed by up to two orders of magnitude relative to solar due to incorporation into dust \citep[e.g.,][]{sav96,wel96,jen09}. 
Ca\,\textsc{ii} absorption studies reveal large line-of-sight variations, reflecting ionization effects and strong depletion \citep{wak01,smo03}. 
The dominance of low-ionization calcium in our X-ray data is therefore consistent with gas-phase calcium tracing only a small fraction of the total elemental budget, while the weak constraints on highly ionized species indicate limited contribution from hot ISM phases \citep{gat18}. 
These results demonstrate the complementarity of X-ray spectroscopy with longer-wavelength studies and the critical role of dust depletion in interpreting calcium abundances. 
The hydrogen column inferred from Ca\,\textsc{i}+Ca\,\textsc{ii}, $N_{\rm H} = 1.19^{+1.22}_{-0.84}\times10^{23}$~cm$^{-2}$, is consistent with that derived from sulfur and argon.  
 
\section{Conclusions and Summary}\label{sec_con}

We have presented a detailed study of X-ray absorption by sulfur, argon, and calcium in the ISM along the line of sight to MAXI~J1744$-$294 using {\it XRISM} Resolve and complementary {\it Chandra} HETG observations. Our main findings are summarized below:

\begin{enumerate}
    \item The S K-edge is dominated by S\,{\sc ii} in the cold ISM and S\,{\sc xiv} in the hot phase. The fraction of sulfur locked in dust is constrained to $<30\%$, consistent with previous X-ray and UV studies \citep[e.g.,][]{jen09,psa24,cor25}. The hydrogen column inferred from S\,{\sc ii} is $N_{\rm H} = 1.14\pm0.13\times10^{23}$~cm$^{-2}$, in broad agreement with other X-ray-based measurements.
    
    \item The Ar K-edge is dominated by Ar\,{\sc ii}, with neutral and highly ionized species constrained by upper limits. These results agree with previous X-ray and optical/UV studies \citep[e.g.,][]{gat24c,sof04,jen09,erz25}, confirming that argon primarily resides in low-ionization phases. The hydrogen column derived from Ar\,{\sc ii} is $N_{\rm H} = 1.34\pm0.32\times10^{23}$~cm$^{-2}$, consistent with the S-based value.
    
    \item For the first time, we report X-ray absorption by Ca\,{\sc i}--Ca\,{\sc iii} in the ISM, enabled by newly computed $R$-matrix photoabsorption cross-sections. Gas-phase calcium is dominated by low-ionization species, while higher-ionization ions are weakly constrained. This is consistent with optical/UV studies showing strong depletion of calcium into dust \citep[e.g.,][]{sav96,wel96,jen09,wak01,smo03}. The hydrogen column inferred from Ca\,{\sc i}+Ca\,{\sc ii} is $N_{\rm H} = 1.19^{+1.22}_{-0.84}\times10^{23}$~cm$^{-2}$, is in good agreement with the S and Ar values.
    
    \item Sulfur and argon, being only weakly depleted, probe multiple ISM phases and yield hydrogen columns representative of the total line-of-sight content. Calcium, heavily depleted into dust, primarily traces the cold and warm neutral ISM. Together, these three elements allow a detailed assessment of ionization structure and dust depletion along the Galactic center sightline.
 
\end{enumerate}

\begin{acknowledgements}  
This work has made use of data from the {\it XRISM} satellite, which is a mission of the Japan Aerospace Exploration Agency in partnership with NASA and ESA. 
We thank the XRISM operation team for accepting our DDT proposal and conducting the observation, along with the XRISM Science Data Center, help desk, and calibration teams, for their continued assistance. 
This work has also made use of software from the HEASARC, which is developed and monitored by the Astrophysics Science Division at NASA/GSFC and the High Energy Astrophysics Division of the Smithsonian Astrophysical Observatory.
This research was carried out on the High-Performance Computing resources of the Raven and Viper clusters at the Max Planck Computing and Data Facility (MPCDF) in Garching, operated by the Max Planck Society (MPG).  

MP acknowledges support from the JSPS Postdoctoral Fellowship for Research in Japan, grant number P24712, as well as the JSPS Grants-in-Aid for Scientific Research-KAKENHI, grant number J24KF0244. Support for SM, KM, and the Columbia University team was provided by \textit{NuSTAR} AO-10 (80NSSC26K0286), \textit{NuSTAR} AO-11 (80NSSC26K0154), \textit{Chandra}\ AO-26 (SAO GO5-26016X) and XMM-\textit{Newton} AO-23  (80NSSC25K0651) programs. 
SM acknowledges support by the National Science Foundation Graduate Research Fellowship under Grant No. DGE 2036197 and the Columbia University Provost Fellows Program.  
MN and HU acknowledge the support by JSPS KAKENHI, Grant Number JP24K00677.
Part of this work was financially supported by Grants-in-Aid for Scientific Research 19K14762, 23K03459, 24H01812 (MS) from the Ministry of Education, Culture, Sports, Science and Technology (MEXT) of Japan. 
\end{acknowledgements}

\bibliographystyle{aa}
\bibliography{my-references}

\end{document}